\documentclass[lettersize,journal]{IEEEtran}
\usepackage{lineno}
\usepackage[utf8]{inputenc}
\usepackage{authblk}
\title{SparseAlign: A Super-Resolution Algorithm for Automatic Marker Localization and Deformation Estimation in Cryo-Electron Tomography}
\author[1,2]{Poulami Somanya Ganguly}
\author[1]{Felix Lucka}
\author[3]{Holger Kohr}
\author[3]{Erik Franken}
\author[2]{Hermen Jan Hupkes}
\author[1,4]{K. Joost Batenburg}
\affil[1]{Computational Imaging, Centrum Wiskunde \& Informatica, Amsterdam, The Netherlands}
\affil[2]{Mathematical Institute, Leiden University, Leiden, The Netherlands}
\affil[3]{Thermo Fisher Scientific, Eindhoven, The Netherlands}
\affil[4]{Leiden Institute of Advanced Computer Science, Leiden University, Leiden, The Netherlands}
\date{}                  
\usepackage{xcolor}
\usepackage{graphicx}
\graphicspath{ {paper_figures/} }
\usepackage{cite}
\usepackage{amsmath}
\usepackage{amsfonts}
\usepackage{amssymb}
\usepackage{mathtools}
\DeclareMathOperator*{\minimize}{minimize}

\usepackage{algorithm}
\usepackage{algorithmic}
\usepackage{caption}
\usepackage{subcaption}
\usepackage{textcomp}

\usepackage[textwidth=2.450cm, textsize=tiny]{todonotes}

\usepackage{xr-hyper}

\makeatletter
\newcommand*{\addFileDependency}[1]{
  \typeout{(#1)}
  \@addtofilelist{#1}
  \IfFileExists{#1}{}{\typeout{No file #1.}}
}
\makeatother

\newcommand*{\myexternaldocument}[1]{%
    \externaldocument{#1}%
    \addFileDependency{#1.tex}%
    \addFileDependency{#1.aux}%
}
\myexternaldocument{supplementary_materials}
\usepackage{hyperref}

\begin{document}

\maketitle

\begin{abstract}
Tilt-series alignment is crucial to obtaining high-resolution reconstructions in cryo-electron tomography.
Beam-induced local deformation of the sample is hard to estimate from the low-contrast sample alone, and often requires fiducial gold bead markers. 
The state-of-the-art approach for deformation estimation uses (semi-)manually labelled marker locations in projection data to fit the parameters of a polynomial deformation model. Manually-labelled marker locations are difficult to obtain when data are noisy or markers overlap in projection data. 
We propose an alternative mathematical approach for simultaneous marker localization and deformation estimation by extending a grid-free super-resolution algorithm first proposed in the context of single-molecule localization microscopy. 
Our approach does not require labelled marker locations; instead, we use an image-based loss where we compare the forward projection of markers with the observed data. 
We equip this marker localization scheme with an additional deformation estimation component and solve for a reduced number of deformation parameters. 
Using extensive numerical studies on marker-only samples, we show that our approach automatically finds markers and reliably estimates sample deformation without labelled marker data.
We further demonstrate the applicability of our approach for a broad range of model mismatch scenarios, including experimental electron tomography data of gold markers on ice.
\end{abstract}

\begin{IEEEkeywords}
Super-resolution imaging, parallel-beam tomography, conditional gradient method, marker-based alignment.
\end{IEEEkeywords}

\section{Introduction} 
Cryo-electron tomography (cryoET) is a powerful imaging technique to resolve the structures of biomolecules and cellular components \textit{in situ}
using an electron microscope \cite{turk2020promise}. In recent years, advancements in detector technology and image processing methods have greatly improved the resolution of structure determination routines
using cryoET, down to near-atomic resolution \cite{KONING201882}.

A typical cryoET workflow consists of tilt-series acquisition, tilt-series alignment and reconstruction, followed by post-processing steps such as per-particle reconstruction refinement, segmentation and sub-tomogram averaging \cite{chen2019complete, pyle2021current}.

\begin{figure*}
    \centering
    \includegraphics[scale=0.55]{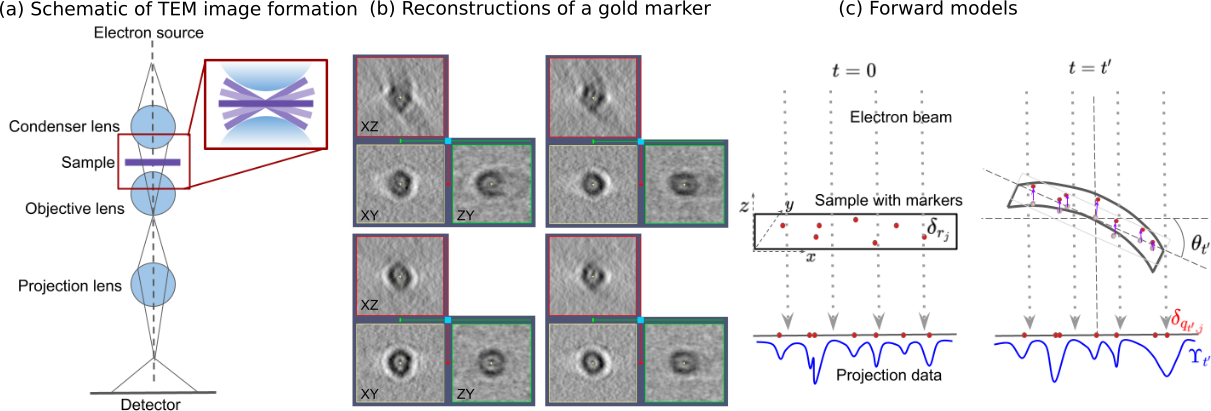}
    \caption{(a) Schematic of image formation using a transmission electron micrscope (TEM). The inset shows sample tilting, which is restricted by the geometry of the microscope. (b) Reconstructions of a gold bead marker using (top two rows) standard alignment without sample deformation compensation and (bottom two rows) with sample deformation compensation. Images reproduced with permission from \cite{fernandez2018cryo}. (c) Forward models used in SparseAlign and the doming model method. At $t=0$ the sample with markers is not deformed. Projected marker locations (red dots) are convolved with a known shape function to yield projection data (blue line). As the sample is tilted, it undergoes doming deformation. At time $t=t^\prime$, the change in marker locations caused by doming (purple upward arrows) leads to a change in the projection data.}
    \label{fig:forward_model}
\end{figure*}

The image formation process in cryoET is as follows. A frozen sample is inserted into a transmission electron microscope (TEM) where it is irradiated with an electron beam, and the resulting transmitted beam lands on the camera to form a TEM image. We show a schematic of TEM image formation in Fig.~\ref{fig:forward_model}(a). For biological samples, the observed image contrast is mainly phase contrast because such samples are made up of light materials and thus are weak scatterers \cite{vulovic2013image}. In contrast, gold markers are strong scatterers and show clear image contrast even under low-dose acquisition conditions. In order to obtain a tomographic \textit{tilt series} (i.e.~a series of projection images for consecutive angles), images of the sample are acquired at different view angles by tilting the sample with respect to the electron beam. 

Aspects of cryoET that distinguish it from other CT setups are as follows. Firstly, the geometry of the experimental system limits the extent to which the sample can be tilted, as shown in the inset of Fig.~\ref{fig:forward_model}(a). As a consequence, projection images can only be acquired for a limited angular range in cryoET,  usually in $[-70^\circ, 70^\circ]$, resulting in a \textit{missing wedge} of information that is not available during reconstruction \cite{bendory2020single}. Secondly, cryoET samples are dose-sensitive, which limits the total dose during acquisition and leads to very noisy projection images when a large number are acquired. Thirdly, the sample undergoes local and global movements during the acquisition procedure, making it difficult to reconstruct with a constant sample assumption. For a detailed discussion on the mathematics of electron tomography we refer the reader to \cite{oktem2015mathematics}. 

The acquired tomographic tilt series must be corrected for global and local sample motion during tilt-series acquisition \cite{amat2010alignment}. Types of global motion include rotations and shifts of the sample with respect to the field-of-view (FoV) captured by the camera. Local motion includes sample deformation induced by the electron beam. When not corrected, sample motion leads to blurred reconstructions and poor resolution of the biological structures extracted by further post-processing \cite{brilot2012beam}. \textit{Tilt-series alignment}, the process of figuring out geometric relationships between projections in the tilt series, provides a way to correct for these effects so that the highest possible resolution can be achieved in subsequent reconstructions. 

Beam-induced local sample deformation is a crucial limiting factor in high-resolution cryoET studies \cite{fernandez2018cryo}. In particular, as shown in Fig.~\ref{fig:forward_model}(b), compensation of local motion during alignment leads to sharper reconstructions and thus more reliable structure determination.
In \cite{fernandez2018cryo}, the authors propose a method to extend currently used alignment methods with a sample deformation term that takes into account local sample motion induced by the electron beam. It has previously been observed that cryoET samples undergo ``doming'' motion, where the sample exhibits an upward deformation perpendicular to the sample plane (Fig.~\ref{fig:forward_model}(c)). The authors of \cite{fernandez2018cryo} model this motion using polynomial surfaces with coefficients that can be estimated as part of a minimization scheme. In addition to global shifts and rotations, the parameters of the doming model are fitted by solving a non-linear least-squares problem.

One of the drawbacks of the doming model approach is that it requires labelled marker locations in the tilt series as input, where the same marker has to be identified in all tilt images such that its locations can be connected to a trace. Markers are usually identified and traced in tilt-series images by template matching, a procedure that is prone to errors when the signal-to-noise ratio in tilt images is low, when markers cluster together or when they overlap in projection while being separate in 3D \cite{amat2010alignment}. 

An additional disadvantage of the doming model method
is the large number of parameters
that must be estimated because no additional prior information on the deformation field is incorporated. 
Without smoothness constraints on the time evolution of the deformation field, the model allows deformation parameters to vary freely over the tilt series and does not penalize unphysical deformations. 

Though not always appropriate, smoothness constraints on local sample motion are reasonable in the context of continuous-tilt cryoET (CTT) data collection, where thousands of very noisy projection images are captured continuously while the stage is tilted with a constant rotation speed \cite{chreifi2019rapid}. This allows for a reduction in the number of doming model parameters.  

We propose extensions to the doming model approach that make it possible to align tilt-series images without labelling markers in the tilt series. Taking inspiration from super-resolution algorithms proposed in the context of single-molecule localization microscopy \cite{boyd2017alternating}, we use a continuous formulation of the marker localization problem, which enables us to formulate an image-based loss and identify marker locations with a localization precision greater than the pixel spacing of the acquired tilt-series data. We equip the localization scheme with an additional deformation estimation routine and solve for the parameters of the doming model. 

In addition, we incorporate a polynomial time dependence of the deformation field, which assumes smoothness of the local sample motion after global motion correction. This assumption helps us reduce the number of deformation parameters by orders of magnitude.
An important aspect of our approach, however, is that it is independent of the choice of deformation field parametrization.

To validate our proposed method, we apply it to simulated data in 2D and 3D as well as experimental data containing gold markers on ice. We study the robustness of our approach with respect to noise, forward model mismatch and deformation model mismatch. We show that we are able to estimate deformation fields and marker locations with similar accuracy as the doming model approach without the need for labelled marker data, and that our method estimates deformation parameters accurately despite model mismatch. 

This paper is structured as follows. In Section \ref{math}, we review the mathematical formulation of the alignment problem and discuss a unifying framework for solving it. We derive the doming model approach in \cite{fernandez2018cryo} as \emph{one} possible choice of alignment method.
We also present the main contribution of our paper: a method that localizes markers and estimates deformation fields without marker labelling. In Section \ref{opt}, we give details of the optimization techniques used to solve our extended problem. In Section \ref{exps}, we describe the numerical experiments performed, and discuss our results on 2D and 3D simulated data as well as experimental data in Section \ref{results}. We end our paper with a critical discussion of our approach and point to possible extensions in Section \ref{conclusion}.

\section{Mathematical Formulation}\label{math}

We consider an initial scanned object $u_0(\rho)$, with $\rho \in \Omega \subset \mathbb{R}^d$ ($d=2,3$ for simulated data and $d=3$ for experimental data), which consists of two distinct components with non-overlapping supports: 
\begin{equation}
    u_0(\rho) = u_0^m(\rho) + u_0^s(\rho),
\end{equation}
where $u_0^m(\rho)$ represents markers and $u_0^s(\rho)$ represents the biological sample. 

This initial object deforms over time, in the sense 
\begin{equation}
    u_t(\rho) = u_0(\rho + D_t(P)(\rho)) =: \big(\mathcal{W}_{D_t(P)} u_0\big)(\rho),
\end{equation}
where $D_t(P,\rho): \mathcal{P} \times \Omega \rightarrow \mathbb{R}^d$ is a time- and space-dependent deformation field parametrized by global parameters $P \in \mathcal{P}$. The action of this deformation field can be represented by a linear warping operator $\mathcal{W}_{D_t(P)}$. The global deformation parameters couple the reconstruction problems for individual markers. Later in this section we discuss appropriate parametrizations for the deformation field.

Projection data $\Psi_t$ of the deforming configuration are generated by applying the continuous Radon transform to $u_t(\rho)$:
\begin{equation}\label{eq:JointImRecDefEst}
    \Psi_t = \mathcal{R}_{\theta_t} u_t(\rho) = \mathcal{R}_{\theta_t} \mathcal{W}_{D_t(P)} \left( u_0^m + u_0^s  \right),
\end{equation}
where $\theta_t$ is the projection angle and the Radon transform for $d=2$ is defined as a line integral over rays: 
\begin{flalign*}
    \mathcal{R}_{\theta_t} [u](s)  &=  \int_{l(s,\theta_t)} u(\rho) \, d\rho \\ l(s,\theta_t) &= \{(x,y) \in \mathbb{R}^2 \,|\, x\cos\theta_t + y\sin\theta_t = s \}.
\end{flalign*}
Projection in 3D for a parallel beam geometry, as in the case for cryoET, can be decomposed into a series of 2D projections \cite{natterer2001mathematics}.

The full tomographic data, obtained over discrete time points $t \in \{t_0,t_1,\ldots,t_T\}$ is a stack of individual projections: 
\begin{equation} \label{eq:JointImRecDefEstStack}
    \Psi := \begin{bmatrix}
    \Psi_0 \\
    \Psi_1 \\
    \ldots \\
    \Psi_T
    \end{bmatrix} = \begin{bmatrix}
    \mathcal{R}_{\theta_1} \mathcal{W}_{D_0(P)} \\
    \mathcal{R}_{\theta_2} \mathcal{W}_{D_1(P)} \\
    \ldots \\
    \mathcal{R}_{\theta_T} \mathcal{W}_{D_T(P)}
    \end{bmatrix}
    (u^m_0 + u^s_0).
\end{equation}

Solving the set of equations \eqref{eq:JointImRecDefEstStack} 
when all the variables - $u_0^m$, $u_0^s$ and $D_t$ - are unknown amounts to solving a joint image reconstruction and alignment problem. 
Most approaches for solving the joint problem alternate between solving \eqref{eq:JointImRecDefEstStack} for one of the three variables while keeping the others fixed. In such schemes, determining a good order for these updates is crucial. 

As markers are designed to have a significantly higher contrast compared to the sample, we can often obtain reasonable first estimates for the marker configuration $u_0^m$ and deformations $D_t$ while ignoring the sample contribution. This corresponds to solving \eqref{eq:JointImRecDefEstStack} by setting $u_0^s = 0$.

One way to parametrize the initial marker configuration $u_0^m$ is to represent it using the continuous locations of markers at $t=0$. Here we represent a single marker as a delta peak at the location of its centre convolved with a fixed, known shape function; the marker configuration is then a 
sum of convolved delta peaks in $\Omega \subset \mathbb{R}^d$:
\begin{equation}
    u_0^m(x) = \sum^M_{j=1} \Big (G \ast \delta_{r_j}(\rho) \Big),
\end{equation}
where $r_j$ are the initial marker locations, $M$ is the total number of markers and $G$ is a known shape function, for instance a Gaussian.

For parallel beam projection, Theorem 1.2 in \cite{natterer2001mathematics} states that:
\begin{equation}\label{eq:forw_proj_one_marker}
   \mathcal{R}_\theta ( G \ast \delta_{r_j}(\rho) ) = \left(\mathcal{R}_\theta G \right) \ast \left(\mathcal{R}_\theta   \delta_{r_j}(\rho)\right) =: G_\theta^p \ast \left(\mathcal{R}_\theta   \delta_{r_j}(\rho)\right).
\end{equation}
Furthermore, the Radon transform of a delta peak is a delta peak in projection space:
\begin{equation}
   \mathcal{R}_\theta   \delta_{r_j}(\rho) =  \delta_{A_\theta r_j}(s),
\end{equation}
where $A_\theta \in \mathbb{R}^{(d-1) \times d}$ is a projection matrix that maps marker locations in configuration space to locations in projection space. We denote the resulting projected marker locations by $q_j := A_\theta r_j$.

We can assume that in contrast to the sample, markers are displaced over time, not deformed. Furthermore, when variations in the global deformation field $D_t$ over the area covered by a marker are small, we can approximate:
\begin{equation}
    \mathcal{W}_{D_t} u_0^m(x) \approx  \sum^M_{j=1} \Big (G \ast \delta_{r_j + D_t(P, r_j)}(\rho) \Big).
\end{equation}
Setting $u_0^s= 0$ and inserting the ansatz above into \eqref{eq:JointImRecDefEst} yields
\begin{multline}\label{eq:MakerLoc1}
    \Psi_t =  \mathcal{R}_{\theta_t} \mathcal{W}_{D_t(P)} u_0^m \approx \sum_{j=1}^M \Big( G_{\theta_t}^{p} \ast  \delta_{A_{\theta_t} (r_j + D_t(P, r_j))} \Big) \\
    = \sum_{j=1}^M \Big ( G_{\theta_t}^{p} \ast  \delta_{q_{t,j}} \Big ),
\end{multline}
where
\begin{equation} \label{eq:MakerLoc2}
    {q}_{t,j} = A_{\theta_t} (r_j + D_t(P, r_j)).
\end{equation}
Using equation \eqref{eq:MakerLoc1} amounts to localizing markers by matching their projection data $\Psi_t \in \mathbb{R}^{(N_\theta \times N_d)}$ (in 2D), where $N_\theta$ is the number of projection angles and $N_d$ is the discretisation of the detector plane. A schematic of this forward model is shown in Fig.~\ref{fig:forward_model}(c), where we indicate 1D projected data with blue lines.

In \cite{fernandez2018cryo}, the authors use projected marker locations over time as the input instead of image data (indicated with red dots in Fig.~\ref{fig:forward_model}(c)) and use the following optimization problem for deformation estimation and marker localization:
\begin{equation}\label{eq:doming_model_opt}
    \minimize_{r_j, P} \quad \sum_{t=0}^{T} \sum_{j=1}^{M} \Big \|\Big ( {\tilde{q}}_{t,j} - A_{\theta_t} (r_j + D_t(P, r_j)) \Big) \Bigg \|^2_2.
\end{equation}
Such an approach assumes that we can identify the projected marker locations $\tilde{q}_{t,j}$
directly, 
despite convolution with $G_{\theta_t}^{p}$. Here and elsewhere, we use symbols with a tilde (e.g.~$\tilde{q}_{t,j}$) to denote measured data and symbols without a tilde (e.g.~$q_{t,j}$) to denote model predictions.

Comparing equations \eqref{eq:MakerLoc1} and \eqref{eq:MakerLoc2}, we find that for each $t$ the dimensions of 2D data for \eqref{eq:MakerLoc2} are $d \times M$ and those of the data for \eqref{eq:MakerLoc1} are $N_\theta \times N_d$. Typical values for $d, M, N_\theta$ and $N_d$ are $3, 20, 100$ and $4096$, respectively, such that $d \times M = 3 \times 20$ and $N_\theta \times N_d = 100 \times 4096$, the latter being approximately $6000$ times the former. Thus, \eqref{eq:MakerLoc2} is a much lower-dimensional problem. Furthermore, the deformation field can be extracted from \eqref{eq:MakerLoc2} in a more direct fashion as it directly describes the corresponding projected marker displacement, not the change in the projection image caused by it. 

However, identifying markers robustly from data is not a trivial problem \cite{amat2010alignment}. It involves solving an optimization problem of the form: $\minimize_{q_{t,j}} \sum_t \| \tilde{\Psi_t} - \sum_j  (G_{\theta_t}^{p} \ast  \delta_{q_{t,j}})\|^2_2$. Marker labelling is generally performed using normalized cross-correlation-based schemes or template matching algorithms. Such methods are error-prone when projection data are noisy or when gold beads are occluded or cluster together in projection data. In such situations, users must manually annotate markers, or manually inspect and correct for incorrect and failed detection in one or more images in the tilt series. This manual intervention leads to time-consuming and subjective labelling. 

To avoid solving the marker identification problem, we take a step back and start directly from \eqref{eq:MakerLoc1}. We solve for marker locations and the deformation field in a least-squares sense. In addition, we do not assume that we know the number of markers beforehand. The resulting optimization problem is as follows: 
\begin{equation}\label{eq:adcg_opt}
    \minimize_{r_j, P, M} \quad \sum_{t=0}^{T} \Big \|\tilde{\Psi}_t  - \sum_{j=1}^M \Big (G_{\theta_t}^{p} \ast  \delta_{A_{\theta_t} (r_j + D_t(P, r_j))} \Big )\Big \|^2_2. 
\end{equation}
The optimization problem above assumes a model for the markers, uses an image-based loss and does not need labelled marker locations like the problem in \eqref{eq:doming_model_opt}. In the following section, we discuss optimisation schemes for solving \eqref{eq:adcg_opt}.

The deformation field $D_t$ can be represented using different basis functions. If one uses localized basis functions, e.g.~the B-spline basis functions often used in non-rigid image registration, one either needs a sufficiently dense sampling of the domain with markers or include suitable regularization constraints \cite{modersitzki2003numerical}. Global basis functions that are supported in the entire domain will only lead to a compact, low-dimensional description of the deformation field with sufficient accuracy if they are chosen based on \textit{a priori} knowledge about the sample deformation. 

In this paper, we use the global basis functions proposed in \cite{fernandez2018cryo}, where the beam-induced sample deformation is modeled with a set of polynomial surfaces. The parametrized sample deformation $ D_t(P, r_j) := [D_{t,x}, D_{t,y}, D_{t,z}]$ is modelled with polynomials in $(x,y,z)$ such that the deformation in each direction is given by
\begin{equation}\label{eq:doming_poly}
    D_{t,k}(r, P) = \sum_{\substack{\alpha, \beta, \gamma \geq 0\\ \alpha + \beta + \gamma \leq d_p}}\Big(P_{\alpha \beta \gamma}(t)\Big)_k x^\gamma y^\beta z^\alpha, \quad k \in \{x,y,z\},
\end{equation}
where $P_{\alpha \beta \gamma}$ are the coefficients of the polynomial and $d_p$ is the degree of the polynomial. In \cite{fernandez2018cryo}, these polynomials are allowed to vary freely over the tilt series, resulting in a large number of free parameters. In 3D, we must estimate $18$ parameters for each tilt for a quadratic deformation model, which amounts to thousands of parameters when the number of tilts is high. One way to reduce the number of parameters, used in \cite{fernandez2018cryo}, is by assuming that the deformation field is constant along the depth ($z$ direction) of the sample. 
with $\frac{(d_p+2)(d_p+1)}{2}$ free parameters. 

To further reduce the number of free parameters, we introduce a temporal dependence in \eqref{eq:doming_poly}, which reduces the number of parameters from $18$ for each tilt to $18$ for the entire tilt series, assuming a quadratic deformation model. Our time-dependent deformation field is given by:
\begin{equation}\label{eq:time_dep_doming_poly}
   D_{t,k}(r, P) = \sum_{\zeta=1}^{d_t} \sum_{\substack{\alpha, \beta, \gamma \geq 0\\ \alpha + \beta + \gamma \leq d_p}} \Big (P_{\alpha \beta \gamma \zeta}\Big)_k x^\alpha y^\beta z^\gamma t^\zeta, t \in [0,1].
\end{equation}
As we reconstruct the first image, there is no way to recover a zeroth order deformation in time. For simplicity, we consider linear time dependence in our experiments, which amounts to setting $d_t = 1$. 

Our method is independent of the choice of parametrization of the deformation field. Other parametrizations, which take advantage of the possible symmetries of the deformation field or additional understanding of the physics underlying the sample behaviour, could also be suitable choices.

\section{Optimization}\label{opt}
In \cite{boyd2017alternating, alberti2019dynamic, ganguly2020atomic}, convex approximations to the minimization problem \eqref{eq:adcg_opt} have been devised by mapping the problem onto the space of measures $\mathcal{M}(\Omega)$. We interpret the marker configuration as a measure $\mu := \sum_{j=1}^{M} w_j \delta_{r_j} \in \mathcal{M}(\Omega)$, where the weights $w_j$ are introduced as a means of relaxing the optimization problem \eqref{eq:adcg_opt}. The weights determine the relative ``importance" of the markers and, as we show later, can be used to remove candidate markers that do not contribute significantly to the data. Mapping the problem to measure space enables us to express the forward operation shown in \eqref{eq:MakerLoc1} in terms of a linear operator, $\Phi_t: \mathcal{M}(\Omega) \rightarrow \mathbb{R}^{N_d}$:
\begin{equation}\label{eq:forw_model_measures}
    \Psi_t = \sum_{j=1}^{M} w_j \Big ( G_{\theta_t}^{p} \ast  \delta_{q_{t,j}} \Big ) =: \Phi_t \mu, \qquad \Psi = \begin{bmatrix}
    \Phi_1 \\
    \Phi_2 \\
    \ldots \\
    \Phi_T
    \end{bmatrix} \mu =: \Phi \mu
\end{equation}

The minimization problem \eqref{eq:adcg_opt} can then be rewritten as the following problem in the space of measures, where the loss is convex in the measure $\mu$:
\begin{equation}\label{eq:adcg_opt_measures}
    \minimize_{\mu \in \mathcal{M}(\Omega)} \quad \ell(\Phi \mu - \tilde{\Psi}), \qquad \ell(\cdot) := \| \cdot \|^2_2
\end{equation}

In \cite{boyd2017alternating}, the authors devised an effective numerical scheme for solving infinite-dimensional convex problems of the type shown above by using a variant of the conditional gradient or Frank-Wolfe method \cite{frank1956algorithm}. They also showed that interleaving the convex Frank-Wolfe iterations with nonconvex local optimization steps improved the convergence of the algorithm. This algorithm, known as the alternating descent conditional gradient (ADCG) method, has been subsequently extended for and applied to a range of application areas \cite{alberti2019dynamic, boyd2017alternating, ganguly2020atomic}.

In this paper, we adapt the ADCG algorithm to solve the marker localization and deformation estimation problems simultaneously. To do this, we perform the Frank-Wolfe iterations as-is but modify the block coordinate descent routine to include an additional deformation estimation step.
At each iteration of the algorithm, we place a new marker at a candidate initial location by solving a linearized approximation of our optimisation problem. Then, we solve a linear optimisation problem to obtain estimates for the weights of all current markers. Local optimisation routines are used to solve for the parameters for the deformation field and to refine the marker support in a bounded region.
Our modified ADCG routine, which we call SparseAlign, is shown in Algorithm \ref{alg:modified_adcg}. Below we describe each step in our method in detail.

\begin{algorithm}[h]
	\caption{SparseAlign}
	\begin{algorithmic}
	\FOR {$n=1:n_{\text{max}}$}
	        \STATE 1) Compute current residual: $\varrho_n \leftarrow \Phi \mu_n - \tilde{\Psi}$
            \STATE 2) Find next marker: ${r}^\ast_{n} \leftarrow \arg\min_{{r} \in \text{grid}} \langle \nabla \ell(\varrho_n), \Psi({r}) \rangle$
            \STATE 3) Update support: $\boldsymbol{r}_{n+1} \leftarrow [{\boldsymbol{r}}_n, {r}_n^\ast]$ 
            \STATE 4) Block coordinate descent:
            
            \STATE\hspace{\algorithmicindent} \textbf{Repeat}:
            \STATE\hspace{\algorithmicindent} (a) Compute weights: \\ 
            \hspace{\algorithmicindent} \hspace{\algorithmicindent} $w_{n+1} \leftarrow \arg \min_w \ell(\Phi \mu_{n+1} - \tilde{\Psi})$
            \STATE\hspace{\algorithmicindent} (b) Prune support: \\
            \hspace{\algorithmicindent} \hspace{\algorithmicindent}$(w_{n+1}, {\boldsymbol{r}}_{n+1}) \leftarrow \texttt{prune}(w_{n+1}, {\boldsymbol{r}}_{n+1})$
            \STATE\hspace{\algorithmicindent} (c) Fit deformation parameters:\\ \hspace{\algorithmicindent} \hspace{\algorithmicindent} ${P}_{n+1} \leftarrow \arg \min_{{P} \in \mathcal{P}} \ell(\Phi \mu_{n+1} - \tilde{\Psi})$
            \STATE\hspace{\algorithmicindent} (d) Improve support:\\  \hspace{\algorithmicindent} \hspace{\algorithmicindent} $\boldsymbol{r}_{n+1} \leftarrow \arg \min_{\boldsymbol{r} \in \mathcal{C}} \ell(\Phi \mu_{n+1} - \tilde{\Psi})$
	\ENDFOR
	\end{algorithmic} 
	\label{alg:modified_adcg}
\end{algorithm}

\paragraph{Adding candidate marker locations}
We use the conditional gradient method to obtain candidate marker locations in steps 2-3. The conditional gradient or Frank-Wolfe method \cite{frank1956algorithm} can be used to solve constrained optimization problems of the type $\minimize_{x\in \mathcal{C}}f(x)$ iteratively, where $C$ is a convex set. The first step in each iteration is to minimize a linearized version of the loss within a specified domain. The linear approximation to a function $f({x})$ at ${x}_k$ is given by
\begin{equation*}
    f_{\text{lin}}({s}) = f({x}_k) + \langle \nabla f({x}_k), {s} - {x}_k \rangle.
\end{equation*}
Minimizing $f_{\text{lin}}({s})$ over a domain $\mathcal{D}_s$ thus amounts to solving
\begin{equation*}
     \minimize_{{s} \in \mathcal{D}_s} \quad \langle \nabla f({x}_k), {s} - {x}_k \rangle.
\end{equation*}

Using our forward model \eqref{eq:forw_model_measures} and the loss function in \eqref{eq:adcg_opt_measures}, we can compute that the linear minimisation step at iteration $n$ is the following optimisation problem over measures $s \in \mathcal{M}_s(\Omega) \subset \mathcal{M}(\Omega)$
\begin{flalign}\label{eq:lmo}
    \minimize_{s \in \mathcal{M}_s(\Omega)} \quad \langle \nabla \ell(\varrho_n), \Phi s \rangle,
\end{flalign}
where $\varrho_n := \Phi \mu_n - \tilde{\Psi}$ is the residual at iteration $n$.

An optimal solution of the above problem is the addition a single new marker with positive weight to the current support of $\mu_n$. This ensures that, at iteration $n$ of the algorithm the measure $\mu$ is supported at $n$ points. Adding only one location at a time has been shown to give the sparsest possible solution \cite{boyd2017alternating}.

Practically, we solve \eqref{eq:lmo} by gridding the domain of marker locations coarsely. The contribution of a single marker at each grid point, $r_{\text{grid}}$, is computed for a current guess of deformation parameters:
\begin{equation*}
    \psi({r_\text{grid}}) = \begin{bmatrix}
    G_{\theta_t}^{p} \ast \delta_{A_{\theta_1}(r_{\text{grid}} + D_1(r_{\text{grid}}))}\\
    G_{\theta_t}^{p} \ast \delta_{A_{\theta_2}(r_{\text{grid}} + D_2(r_{\text{grid}}))}\\
    \ldots \\
    G_{\theta_t}^{p} \ast \delta_{A_{\theta_T}(r_{\text{grid}} + D_T(r_{\text{grid}}))}\\
    \end{bmatrix}
\end{equation*}
Then, the inner product of the current residual with the forward projection of a marker located at each grid location is calculated. The grid location ${r}^\ast_{\text{grid}}$ with the smallest inner product with the residual is chosen as the next candidate location:
\begin{equation}\label{eq:lmo_opt}
    {r}^\ast_{\text{grid}} = \arg\min_{{r} \in \text{grid}} \quad \langle \nabla \ell(\varrho_n), \psi({r}) \rangle.
\end{equation}

\paragraph{Optimizing weights}
Once we have optimized for marker locations, we can optimize the weights of each marker as shown in steps 4(a)-(b). Note that the model \eqref{eq:forw_model_measures} depends linearly on the weights $w_j$, $j \in \{1,2,\ldots,M\}$. Thus, with the number of markers, marker locations and deformation parameters fixed, the weights $w_j$ can be estimated by solving the following linear least-squares problem
\begin{equation}
    \minimize_{w \in [0,1]^{n}} \quad \|\ell(\Phi \mu_n - \tilde{\Psi})\|^2_2.
\end{equation}

All weights $w_j$ are constrained to lie in $[0,1]$ and represent the relative importance of marker contributions to the data. Markers with weights close to zero can be removed by an additional \texttt{prune} routine that removes all markers with a weight lower than a predefined threshold. In some cases an additional \texttt{prune} routine can be used to remove markers with small weights at the end of a full algorithm run. This further ensures that the solution obtained is the sparsest possible marker configuration required to explain the data $\tilde{\Psi}$.

\paragraph{Refining initial marker locations} At each iteration, we perform the nonconvex local optimization step shown in 4(d) to refine our estimates for the initial marker locations. This step was first proposed in \cite{boyd2017alternating} as a way to speed up convergence of the conditional gradient method. 

Refining the support of the current measure $\mu_n$ without changing the number of markers ensures that markers are moved off the grid locations used in steps 2-3. It also imparts some of the rapid local convergence qualities of nonconvex optimisation \cite{boyd2017alternating}. In our implementation, we use the L-BFGS-B algorithm to perform local optimisation over initial marker locations.

\paragraph{Estimating deformation parameters}
The optimization problem behind step 4(c) is given by
\begin{equation}\label{eq:DeformEst}
\minimize_{{P} \in \mathcal{P}} \quad \sum_{t=0}^{T} \Big \|\tilde{\Psi}_t  - \sum_{j=1}^M w_j \Big (G_{\theta_t}^{p} \ast  \delta_{A_{\theta_t} (r_j + D_t(r_j, P))} \Big )\Big \|^2_2 ,
\end{equation}
which is a difficult nonconvex problem that is often studied in the context of image correspondence problems such as image registration or optical flow estimation \cite{evangelidis2008parametric}. We use L-BFGS-B initialized at the current $P_n$ to compute a local update $P_{n+1}$ for the parameters of the deformation field. 

\paragraph{Coarse-to-fine scheme for large data}

One of the challenges of solving \eqref{eq:DeformEst} is that the objective function is flat if the forward projection of the current marker configuration and the data do not share the same support, and gradient-based optimization schemes such as L-BFGS-B have a hard time locating a minima. This easily happens for small objects, such as markers, embedded in large projection images. The remedy is typically to smooth both images with a Gaussian, compute a deformation field on the smoothed problem, and use the solution of the smoothed problem to initialize the optimization of the original problem. 

Gaussian smoothing followed by downsampling removes high image frequencies and one starts matching only the low frequencies. 
For noisy data, downsampling has the additional advantage of denoising the data. Furthermore, for large experimental data, where each tilt image has pixel dimensions $ 4096 \times 4096$, warm-starting the optimization at high resolutions with good initial values ensures that not many expensive iterations have to be performed.

For realistic simulation data and experimental data, we use a coarse-to-fine scheme where the marker localization and deformation estimation problem is solved at successively finer resolutions using the results at the coarser resolutions as initialization.

At full resolution, we generate the forward projection of a single marker using \eqref{eq:forw_proj_one_marker} followed by sampling on a spatial grid $X_f$ with $N_d$ grid points. Thus, the discretized forward projection of the full marker configuration can be written as
\begin{equation}
    \Psi_t = \sum_j w_j S^f \mathcal{G}_{(q_{t,j}, \tau_f)},
\end{equation}
where $S^f$ is the sampling operator associated with the spatial grid $X_f$ and $\mathcal{G}_{(q_{t,j}, \tau_f)}$ is a Gaussian centred at $q_{t,j}$ with standard deviation $\tau_f$. 

For obtaining measured data at coarse resolutions, we downsampled the full-resolution measured data $\tilde{\Psi}_t$ at each time after Gaussian convolution to prevent aliasing artefacts \cite{harris2021multirate}. Thus, the coarse-resolution data were given by $\tilde{\Psi}_t^c := \mathcal{H}^c (\mathcal{G}_{\tau_a} \ast \tilde{\Psi}_t)$, where $\mathcal{H}^c$ is a downsampling operator associated with a coarse grid $X_c$ and $\mathcal{G}_{\tau_a}$ is an anti-aliasing Gaussian. For integer downsampling factors $\eta := |X_c| / |X_f|$, $\mathcal{H}^c$ only keeps pixels separated by $\eta$ in the coarse-resolution image.

We approximated matching forward projection data $\Psi^c_t$ directly from marker locations using our forward model \eqref{eq:MakerLoc1} by sampling the Gaussian-convolved projected marker locations on the coarse grid $X_c$:
\begin{equation}\label{eq:scaled_forw_proj}
    \Psi_t^c = \sum_j w_j S^c \mathcal{G}_{(q_{t,j}, \tau_f)},
\end{equation}
where $S^c$ is the sampling operator associated with the coarse grid $X_c$.

\section{Numerical Experiments}\label{exps}
In this section we describe our experiments with simulated and real data. Implementation notes with details of software packages used are provided in Section \ref{impl-notes} of the Supplementary Materials. 

\subsection{Illustrative 2D example}

\paragraph{Ground truth}
We used a simple simulated sample to elucidate properties of our algorithm in 2D. The FoV was taken to be $[-L/2, L/2]$ along both axes, with the canonical length scale $L=1$. The ground truth sample consisted of 10 gold bead markers confined to a thin rectangular region: $x \in [-2L/5,2L/5], z \in [-L/10, L/10]$. We chose this geometry for our 2D sample to mimic the geometry of experimental cryoET samples. 

For simplicity, we considered deformation field components to be zero along the horizontal ($x$) direction. In the vertical ($z$) direction, we assumed the deformation to be given by a quadratic polynomial of $x$ and $z$:
\begin{equation}\label{eq:ground_truth_defo}
    D_{t,z}(r,P) = (P_{0} + P_{1}x + P_{2}z + P_{3}x^2 + P_{4}z^2 + P_{5}xz)t =: D_{1,z}t,
\end{equation}
with $P_0 = 0\ L$, $P_1 = P_2 = -1$, $P_3 = P_4 = P_5=-1\ L^{-1}$, and $t$ taking values in $[0, 1]$ 

\paragraph{Projection data} We generated projection data using the forward model in \eqref{eq:forw_model_measures} over a set of discrete projection angles $\theta \in [-70^\circ, 70^\circ)$, $N_\theta = 20$. Practically, we computed the continuous Radon transform of each marker, followed by a continuous 1D Gaussian convolution in projection space. The Gaussian-convolved projection was then discretized on a detector grid with $N_d = 64$.
At each projection angle, the projection was then a 1D profile. All the projections were rearranged in a sinogram with dimensions $N_\theta \times N_d$.

For comparison, we also generated input data for the doming model method in \cite{fernandez2018cryo}. These data were the projected locations of each marker over the same series of projection angles. 

\subsection{Simulated 3D examples}

\paragraph{Ground truth} We used a 3D configuration of markers to test the robustness of our method to noise and to mismatches in the forward model. We used 20 randomly placed 
markers in a thin region in 3D with dimensions $819.2\ \text{nm} \times 819.2\ \text{nm} \times 100.0\ \text{nm}$. The sample used was the same as that described in \ref{tem_sims}. 

We considered deformation field components to be non-zero only along the $z$ direction; this component was then given by:
\begin{equation}
    D_z(x, y, z, t) = (P_0 + P_1 x^2 + P_2 y^2) t,
\end{equation}
with $P_0 = 200\ \text{nm}$, $P_1 = P_2 = -100\ \text{nm}^{-1}$, and $t$ taking values in $[0, 1]$.

\paragraph{Projection data} We generated projection data along 140 equispaced projection angles in $[-70^\circ, 70^\circ]$ using a Gaussian with standard deviation $15 \text{nm}$ as the shape function of individual markers. Each projection image was discretized on a $64 \times 64$ pixel grid.

To convert the intensities in these generated images to meaningful electron counts, we used that the expected electron count in any pixel is given by $I = I_0 e^{-V_{\text{abs}}C \times \delta x}$, where $I_0$ is the incoming electron count, $V_{\text{abs}}$ is the absorption potential of gold nanoparticles ($5.39 V$ for a $300 keV$ electron beam, treating the gold as amorphous), $C$ is the interaction constant ($0.00653 V^{-1}nm^{-1}$ at $300 keV$) and $\delta x$ is the path length travelled by electrons through a gold marker. This path length is equal to the product of the diameter of the gold bead, which we take to be $15 \text{nm}$, and the intensity in our generated images. For our experiments, we generated data with $I_0 = 2^n, n \in \{6, 7, 8, 10, 12, 14\}$.

\paragraph{Gaussian noise} To test the properties of our approach for noisy data, we performed experiments with data corrupted with additive Gaussian noise, such that
\begin{equation*}
    \Psi_{\text{noisy}} = \Psi_{\text{clean}} + \mathcal{N}(0, \sigma^2_{\text{noise}}),
\end{equation*}
where $\Psi_{\text{clean}}$ are the data scaled to physical electron counts and $\sigma_{\text{noise}}^2$ is the variance of the noise added.

We performed experiments using $\sigma^2 = 2^n$, $n \in \{7, 8, 10, 12, 14\}$. For each noise setting, $10$ independent experiments were performed and the results were averaged to obtain mean values for the metrics. Each independent experiment was initialized with a with a different random seed.

\paragraph{Poisson noise} We also generated a series of Poisson noise-corrupted data by varying the electron count per pixel per frame, $I_0$. For $I_0 = 2^n$, $n \in \{6, 8, 10, 12, 13, 14\}$, we generated Poisson-distributed electron counts at each pixel using:
\begin{equation}
    \Psi_{\text{noisy}} = \mathrm{Poi}(\Psi_{\text{clean}}),
\end{equation}
where $\Psi_{\text{clean}}$ are the data scaled to physical electron counts and $\mathrm{Poi}(\cdot)$ denotes a Poisson random variable. The Poisson-noise data were generated to have comparable signal-to-noise ratios as those of the Gaussian-noise data. For each noise instance, we performed 10 independent experiments with different random seeds and averaged over the obtained metrics.

\subsection{Realistic TEM simulations}\label{tem_sims}

We used the TEM-simulator software \cite{rullgaard2011simulation} to generate physically plausible simulations of TEM images from a specification of a 3D sample. To simplify matters, the sample consisted purely of gold particles in vacuum, thus disregarding the ice buffer and other sample structures. The purpose of this numerical experiment was to test our algorithm in situations where its forward model did not match the one used for data generation. In particular, the explicit assumption of Gaussian shape of gold particles and the implicit assumption of additive uncorrelated noise characteristics were violated.

The test sample consisted of 20 gold particles of 15nm diameter, randomly distributed in a slab of dimensions $819.2 \text{nm} \times 819.2 \text{nm} \times 100.0 \text{nm}$ in $x, y, z$ space. Over time, this sample was simulated to undergo a deformation described by the vector field
\begin{equation}\label{eq:TEM_sim_gt_defo}
    D_z(x, y, z, t) = (P_0 + P_1 x^2 + P_2 y^2) t, \quad D_x = D_y \equiv 0
\end{equation}
with $P_0 = 200\ \text{nm}$, $P_1 = P_2 = -100\ \text{nm}^{-1}$, and $t$ taking values in $[0, 1]$. This amount of deformation (200 nm at $x=y=0,\ t=1$) is an exaggerated version of a doming motion observed in practice. The large amplitude was chosen to make the effects under investigation easier to observe.

Assuming constant tilt speed, the time $t$ was mapped to a tilt angle $\theta$ according to $\theta_i = -70^\circ + t_i \cdot 140^\circ,\ t_i = \frac{i}{140}\,, i=0,\dots,140$. At each tilt angle, a projection image was simulated according to the weak phase object approximation model \cite{vulovic2013image}, taking the contrast transfer function (CTF) of the optical system into account (see \cite{rullgaard2011simulation} for details). We used electrostatic potential values of $V=0$ for vacuum and $V = (29.87 + \imath\cdot 5.39)$ Volt for (amorphous) gold. The CTF parameters were chosen as  $\varDelta z = 8$ $\mu$m (defocus), $C_C = 2$ mm (chromatic aberration) and $C_S = 2.7$ mm (spherical aberration).

The size of each projection image was chosen equal to the $x-y$ dimensions of the sample, subdivided into $(N_x, N_y) = (512, 512)$ pixels, each of size $1.6$ nm.

\paragraph{Noiseless data}

The noiseless images generated by TEM-Simulator correspond to probability densities of detecting an electron at a given location in the detector plane. Therefore, scaling with the average number of incoming electrons per pixel area results in each pixel value representing the expected number of electrons measured in that pixel, also referred to as ``infinite dose" case.


\paragraph{Noise generation}

In a real experiment, a finite number of electrons interacts with the sample and is detected at the camera. This process was modeled with a Poisson random variable $\mathrm{Poi}(\lambda_k)$ per pixel, where the parameter $\lambda_k = I_0 \Psi_k$ equals the intensity of the $k-$th pixel in the scaled noiseless image. This noise model applies to a perfect counting camera. However, cameras operating in integration mode have a nontrivial point spread function because charge from one incident electron can leak into neighboring pixels, triggering multiple detection events. Furthermore, signal and noise transfer vary with spatial frequency. These two effects are characterized by the MTF (modulation transfer function) and DQE (detective quantum efficiency) of the camera and lead to a slight image blur and, more importantly, correlated noise \cite{vulovic2013image}. The noisy images in these numerical experiments made use of this model.


\paragraph{Pre-processing for noisy data}

For data with correlated Poisson noise, we performed the following pre-processing steps. First, we used noiseless data to perform segmentation with Otsu's method \cite{otsu1979threshold}. We obtained a mask for the markers in the tilt series from this segmentation procedure, which we used to compute average background and marker intensities in the noisy tilt series. Second, we shifted the range of the noisy data by subtracting its minimum value and applied the Anscombe transform to our shifted data. The Anscombe transform 
\begin{equation*}
    \text{Anscombe}(\tilde{\Psi}) := 2 \sqrt{\tilde{\Psi} + 3/8}
\end{equation*}
was used as a variance-stabilizing transformation to obtain data with an approximately constant standard deviation \cite{anscombe1948transformation}. Finally, we subtracted the average background intensity and divided by the average bead intensity in the data. 

\subsection{Experimental data} 
For our experimental data we did not aim at a biologically relevant sample, but instead at a sample with gold beads as the only prominent features. We deposited 20nm gold particles on a lacey carbon grid, which was plunge-frozen in liquid ethane using a Thermo Scientific Vitrobot.

We acquired a tomographic tilt series using the Thermo Scientific Tomography 5.5 software package on a Thermo Scientific Titan Krios electron microscope equipped with a Thermo Scientific Falcon 3EC camera. An area in a hole with 15 gold beads was selected. A magnification of 37000x was chosen for a pixel size of $1.949\AA$ and a field of view of 800~nm.  The sample was tilted from -60 to +60~degrees with a tilt step of 2~degrees. Each image in the tilt series had an electron dose of 0.198 $e^{-}/\AA^{2}$.

\paragraph{Cross-correlation-based global alignment} 
Projection images were globally shift-aligned using the cross-correlation-based routine in Thermo Scientific Inspect3D. 

\paragraph{Data pre-processing}
Not all projections were globally aligned correctly using the cross-correlation-based alignment routine. We inspected the tilt series visually for any misaligned projections and removed these. 
This resulted in a total of 27 projections that were then used for estimating local sample deformation.
Next, we deleted 256 pixels from each of the four borders of the tilt series images to get rid of missing image data added by the cross-correlation-based alignment routine. Only one marker, near the top edge of the tilt series images, was discarded because of edge removal.
After applying the Anscombe transform, we subtracted the mean of the tilt series; because most pixels were background pixels, this ensured that the average background intensity was close to 0.
Finally, all tilt series pixels were divided by the average marker intensity to ensure that, in accordance with our forward model, the markers had an average intensity of approximately 1. To determine the average bead intensity in experimental data, we inspected the tilt series visually and used the average intensity in three small square regions around three beads.

\subsection{Evaluation criteria}

To quantify the accuracy of our estimated deformation fields with respect to the ground truth, where available, we used the following evaluation criteria.
First, the estimated and ground truth deformation parameters were used to compute the deformation field at $t=1$ on a gridded FoV of dimensions $1000 \times 1000$ (for 2D) and $1000 \times 1000 \times 1000$ (for 3D), using equation \eqref{eq:ground_truth_defo}.
Next, the vectorial difference between estimated and ground truth deformation fields at $t=1$ was computed at each grid point:
\begin{equation}\label{eq:defo_est_error}
    E(r_{\text{grid}}) = \|D_{1,z}^{\text{gt}}(r_{\text{grid}}) - D_{1,z}^{\text{est}}(r_{\text{grid}})\|^2_2
\end{equation}
This deformation estimation error was averaged over the whole grid to obtain the global deformation estimation error and averaged only at the ground-truth marker locations to obtain the deformation estimation error at markers:
\begin{equation}\label{eq:mean_defo_est_error_global}
    E_{\text{global}} = \dfrac{1}{N_{\text{grid}}} \sum_{\text{grid}} E(r_{\text{grid}})
    \end{equation}
    \begin{equation}\label{eq:mean_defo_est_error_local}
    E_{\text{markers}} = \dfrac{1}{M} \sum_{j=1}^{M} E(r_{j})
\end{equation}
where $N_{\text{grid}} = 10^9$ for 3D and $N_{\text{grid}} = 10^6$ for 2D. 

\section{Results}\label{results}
\begin{figure*}
    \centering
    \includegraphics[scale=0.6]{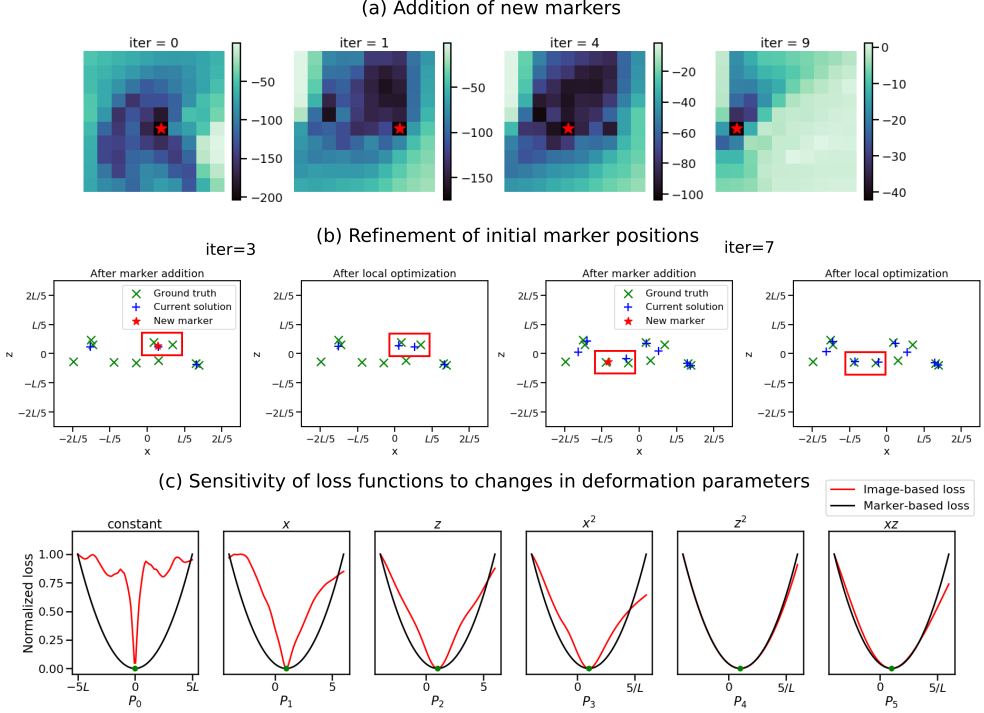}
    \caption{Three steps in SparseAlign. (a) Addition of new markers is performed on a coarse grid using the optimisation problem \eqref{eq:lmo_opt}. The grid location with the smallest pixel intensity in the heatmap is chosen as the next candidate location, which is indicated with a red star. (b) Refinement of initial marker locations is performed using L-BFGS-B. The two leftmost plots show one step of marker addition followed by local optimization; the two rightmost plots show another step of marker addition and local optimization. In both cases, after addition of a new marker (red star), local optimization ensures that all current markers (blue plus signs) are brought closer to the ground truth locations (green crosses). We indicate the areas where this improvement is clearest with red rectangles. (c) Sensitivity of the marker-based loss (black line) used in the doming model approach and our image-based loss (red line) to changes in deformation parameter values. For each plot, the loss was normalized independently with respect to its maximum value.}
    \label{fig:algo_props}
\end{figure*}

\paragraph{SparseAlign adds markers with small displacements first}
In Fig.~\ref{fig:algo_props}(a) and (b), we show how SparseAlign localizes markers. At each iteration, markers are added by solving the linearized problem \eqref{eq:lmo_opt} on a coarse grid. We show the values of the objective function at each grid location in Fig.~\ref{fig:algo_props}(a). The first marker added is a marker close to the centre of the field of view, where the displacement of markers is smallest. This corresponds with the fact that all deformation parameters are set to zero for the first iteration. After the first iteration, when we start optimizing for the deformation parameters, markers that show larger displacements are added. In Fig.~\ref{fig:algo_props}(b), we show two examples of marker location refinement. The two plots on the left show marker addition and refinement at iteration 3; a new marker, indicated with a red star, is added at a grid location. Local optimization then allows us to move this marker as well as all currently placed markers (blue plus signs) off the grid and closer to the ground truth locations (green crosses). The two plots on the right show another step of marker addition and local optimization at iteration 7. In both cases, local optimization helps to improve the solution close to the region where the new marker is added. We indicate this region with a red rectangle in the plots.

\paragraph{SparseAlign's image-based loss is not convex with respect to deformation parameters}
In Fig.~\ref{fig:algo_props}(c), we plot the image-based loss in \eqref{eq:adcg_opt} as a function of each deformation parameter separately, while holding other parameters and marker locations fixed at their respective ground truth values. For comparison we also plot the marker-based loss in \eqref{eq:doming_model_opt}. 
Finally, each plot is normalized with a different normalization constant, equal to the maximum value of the loss for that parameter. 
For each parameter, the marker-based loss is a near-perfect quadratic function with a minimum at the ground truth parameter value. The image-based loss function shares the same minima but differs from the marker-based loss at higher parameter values. In general, the image-based loss function is only convex in a small region around the global minimum. As we move away from the minimum, the loss function increases for each parameter until, at large parameter values, markers move out of the field of view and the loss shows other minima (as in the plot for $P_0$) or flattens and dips (as in the plots for $P_1$ through $P_3$). Gradient-based schemes can thus get caught in local minima if parameter values are very far away from the true minimum at initialization.

\begin{figure*}
    \centering
    \includegraphics[scale=0.6]{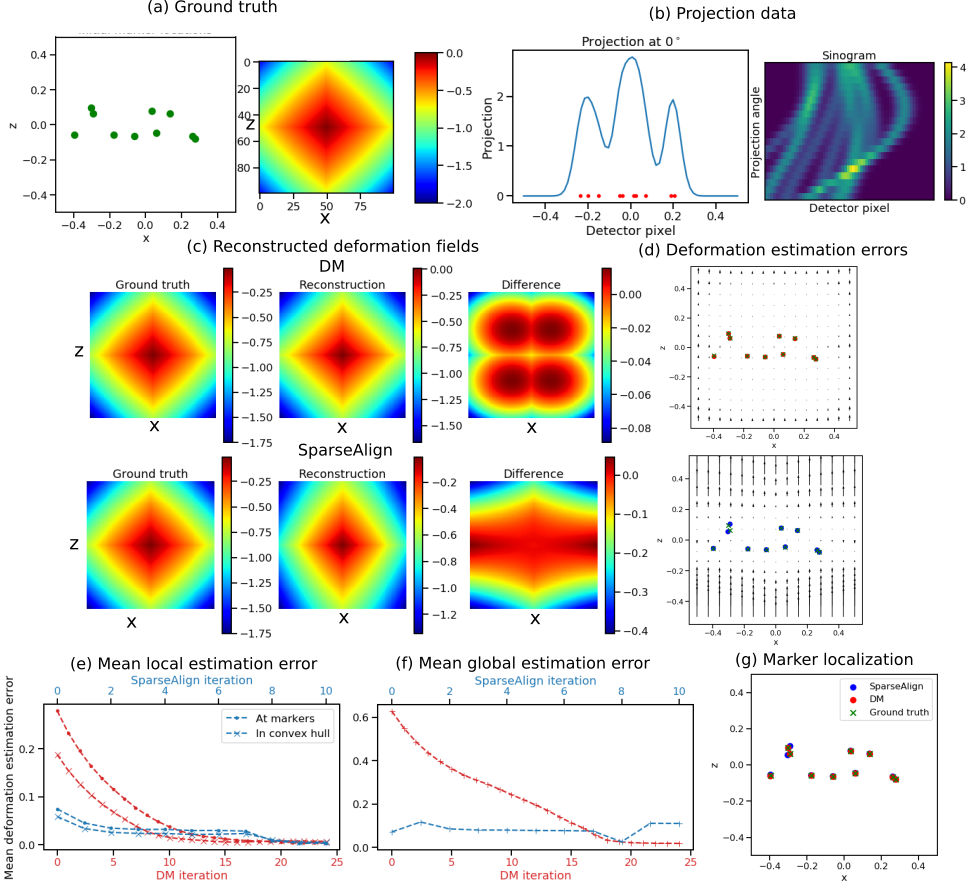}
    \caption{Marker localization and deformation estimation using SpaseAlign and the doming model method (DM). (a) Ground truth initial marker locations and deformation field component along the $z$-axis at $t=t_1$, $D_{1,z}$. (b) Input data for DM are the projected marker locations indicated with red dots. Projection data for SparseAlign at $0^\circ$ is a 1D profile that is a superposition of Gaussians; we indicate this data in blue. The full sinogram data is a stack of projections taken along tilt angles in [-$60^\circ$, $60^\circ$). (c) Reconstructed deformation fields using DM and SparseAlign. In both cases, errors are largest at the boundaries of the field of view (FoV), where no markers are present. (d) Deformation estimation error \eqref{eq:defo_est_error} obtained using DM and SparseAlign. Errors are comparable in the convex hull of markers; errors outside the convex hull are larger when using SparseAlign. (e)-(f) Mean local and global deformation estimation errors \eqref{eq:mean_defo_est_error_global}-\eqref{eq:mean_defo_est_error_local} as a function of DM and SparseAlign iterations. (g) Localized initial marker locations using SparseAlign (blue circles) and DM (red circles) overlaid with the ground truth marker locations (green crosses).}
    \label{fig:simple_2D}
\end{figure*}

\paragraph{SparseAlign estimates deformation parameters with an accuracy comparable to that of the doming model}

In Fig.~\ref{fig:simple_2D} we illustrate the differences between the doming model optimization used in \cite{fernandez2018cryo}~and our method. We use the simple 2D sample shown in Fig.~\ref{fig:algo_props} with a quadratic deformation field along the vertical ($z$) direction. 

Input data for the doming model (`DM') optimization are indicated with red dots in Fig.~\ref{fig:simple_2D}(b); projection data for SparseAlign is a 1D profile indicated with a blue line. The set of line profiles can be rearranged to give a sinogram for the SparseAlign data. 

In Fig.~\ref{fig:simple_2D}(c), we show the reconstructed deformation fields obtained using the two methods. In Fig.~\ref{fig:simple_2D}(d), we illustrate the vectorial deformation field error \eqref{eq:defo_est_error} in both cases. We observe that the error in
the convex hull of the markers is comparable using both methods. This is true despite the fact that our method does not need labelled marker locations and minimizes a more complicated image-based loss function. In regions without markers, our method shows larger errors. This is an indication of the greater ill-posedness of our deformation estimation problem \eqref{eq:DeformEst}. 

In Fig.~\ref{fig:simple_2D}(e-f), we compare mean deformation estimation errors \eqref{eq:mean_defo_est_error_local} and \eqref{eq:mean_defo_est_error_global} for both methods at the ground truth marker locations and in the entire FoV. Mean deformation estimation errors at marker locations are comparable for both methods although the global mean error is higher for SparseAlign. The larger global error, however, is not significant because the major contribution comes from boundaries where no sample is present. Marker localization using SparseAlign and DM gives comparable results, as illustrated in Fig.~\ref{fig:simple_2D}(g).

\paragraph{Deformation estimation accuracy reduces almost linearly for additive Gaussian noise}
In Fig.~\ref{fig:effect_of_noise}, we show results on 3D simulated data. The ground truth marker configuration and deformation field are shown in Fig.~\ref{fig:effect_of_noise}(a). We used different noise settings to probe the properties of our method for data corrupted with Gaussian and Poisson noise. The mean deformation estimation error plots for Gaussian noise show an almost linear decrease in deformation estimation accuracy for increasing Gaussian noise. 

The dependence of deformation estimation error on noise is more complicated in the case of Poisson noise. As shown in the plots in Fig.~\ref{fig:effect_of_noise}(c), we do not see a linear dependence as in the case of Gaussian noise. The difference in accuracy between deformation estimation results for low and high electron counts is also smaller. This suggests that the mismatch between Poisson noise data and data generated from our forward model is greater than the mismatch in the case of comparable Gaussian noise.

\paragraph{Model mismatch does not affect deformation estimation significantly}
We used physically plausible TEM simulations to generate data where the forward model of SparseAlign did not match the data generation model.

\begin{figure*}
    \centering
    \includegraphics[scale=0.42]{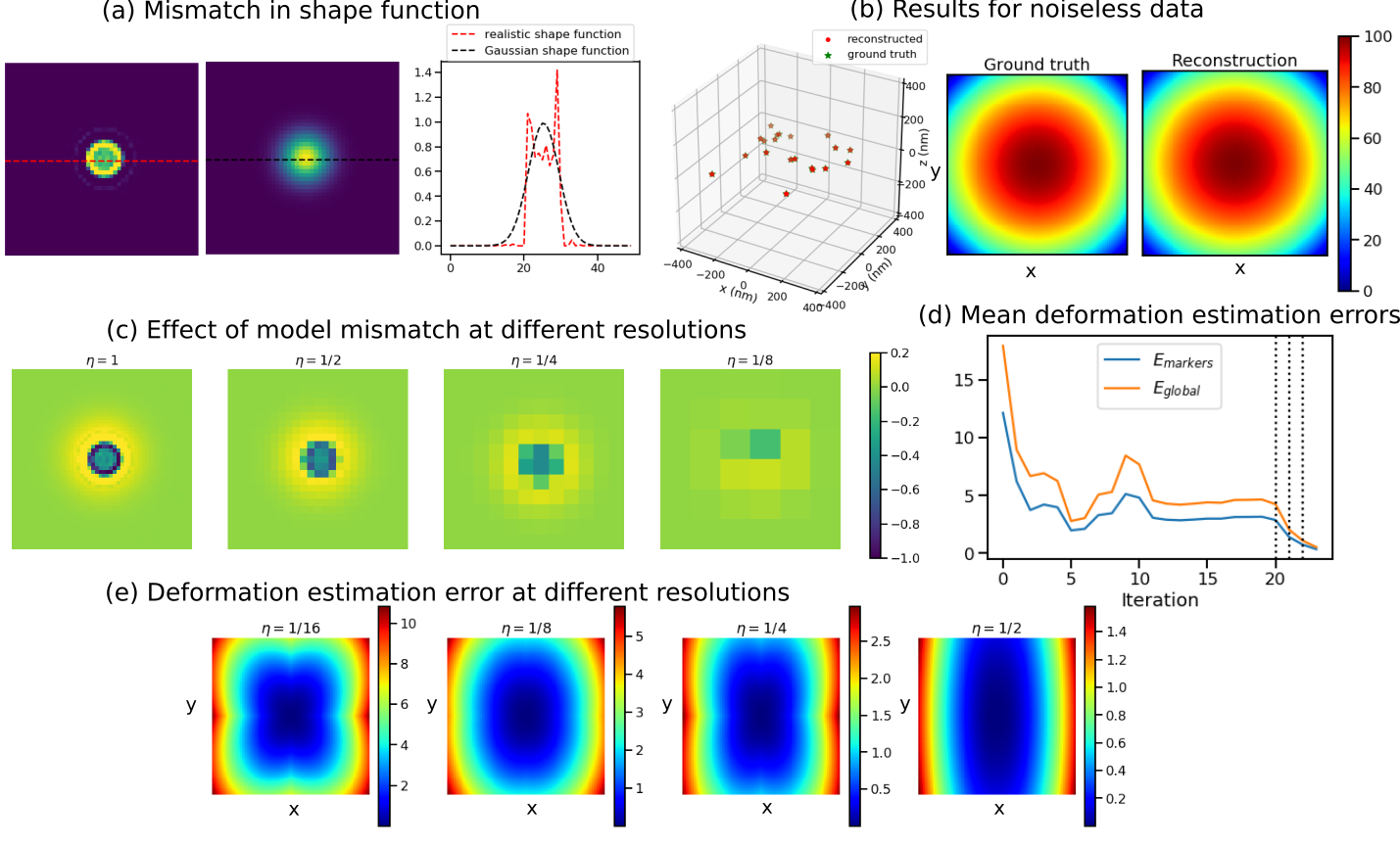}
    \caption{(a) Mismatch in shape function. (left) 2D projection of a single marker generated using the TEM simulator. (centre) Projection of a Gaussian marker used in our forward model. (right) Profiles of both shape functions.
    (b) Marker localization results (left) and deformation estimation results in nm (right) for noiseless realistic data.
    (c) Difference between forward projected marker locations and observed data (a small region around a single marker is shown). The difference due to model mismatch is largest at the fine resolutions.
    (d) Mean deformation estimation error at ground truth marker locations and in the entire FoV for different iterations. Resolution changes in the coarse-to-fine scheme are indicated with black dotted lines.
    (d) Absolute error of estimated deformation field with respect to the ground truth at different values of the downsampling factor $\eta$.}
    \label{fig:model_mismatch}
\end{figure*}

In these data, the shape function of a gold bead marker is not a Gaussian. In Fig.~\ref{fig:model_mismatch}(a), we show the profile of a marker in projection data generated using the TEM-simulator package \cite{rullgaard2011simulation} and the profile of a marker using our forward model. We assumed that the size of gold bead markers and the pixel size of projection images are known, so that the width of the Gaussian can be computed.

In Fig.~\ref{fig:model_mismatch}(b), we show results on marker localization and deformation estimation using noiseless data. The ground truth marker configuration and deformation field are the same as those shown in Fig.~\ref{fig:effect_of_noise}(a). The results we show in Fig.~\ref{fig:model_mismatch}(b) are those obtained at the final step of a coarse-to-fine scheme, where we solved for marker localizations and deformation parameters at increasing resolutions using downsampling factors $\eta = 1/16, 1/8, 1/4, 1/2$. The final result of such a scheme shows a good qualitative match between reconstructed and ground truth marker locations and deformation fields. We stopped at $\eta = 1/2$ because the effect of model mismatch, which we discuss in the next paragraph, is greatest at high resolutions. Moreover, our current implementation is unable to handle very large data sizes, an area we plan to improve in a future work.

In Fig.~\ref{fig:model_mismatch}(c), we show the effect of model mismatch at different resolutions using plots of the difference between our forward projected reconstructed markers and the observed data. We see that the effect of model mismatch is most pronounced at the finest resolutions. This indicates why using a coarse-to-fine scheme, where we obtain initial guesses for marker locations and deformation parameters by solving the problem in a coarse resolution first, leads to reasonable results despite the difference in forward models. 

We plot mean deformation estimation errors \eqref{eq:mean_defo_est_error_local} and \eqref{eq:mean_defo_est_error_global} for each iteration in Fig.~\ref{fig:model_mismatch}(d). Jumps in resolution are indicated with dotted lines. Here we observe that the maximum reduction in deformation estimation error is achieved at the coarsest resolution. The initial guesses obtained are then refined subsequently at each finer resolution. The stopping criterion we used to jump to a higher resolution was to check whether the absolute difference in loss at each new iteration was greater than a pre-set tolerance value (here, $10^{-6}$). 

Finally, in Fig.~\ref{fig:model_mismatch}(e), we illustrate the deformation estimation error \eqref{eq:defo_est_error} at each resolution. Here we observe that, at the coarsest resolution, the error is already small near the centre of the FoV, where a majority of markers is present. At higher resolutions, the refinement in deformation parameters ensures smaller errors at the boundaries and indicates improvements in the values of estimated parameters. 


\paragraph{Marker localization is poor for data with correlated Poisson noise} In Fig.~\ref{fig:correlated_noise}, we show results of our method on data with realistic markers and realistic correlated noise using the ground truth marker configuration and deformation field in Fig.~\ref{fig:effect_of_noise}.

We observe that marker localization for correlated noise-corrupted data is poorer than that for noiseless data (shown in Fig.~\ref{fig:model_mismatch}). At the end of a coarse-to-fine scheme, two markers are not localized and a few markers with small weights are added to the reconstruction domain. These small weighted markers were removed with a further thresholding step, where markers with weights less than $0.1$ were discarded. Improving marker localization might need changes to the forward model used, an aspect that needs further research; however, in our experiments, marker localization did not have a significant effect on deformation estimation accuracy, as seen from the reconstructed deformation field shown in  Fig.~\ref{fig:correlated_noise}(a). 

In Fig.~\ref{fig:correlated_noise}(b), we show plots of mean deformation estimation errors. Note that the same stopping criterion as that used for noiseless data ensured that more iterations were performed at finer resolutions for data with realistic noise.  

In Fig.~\ref{fig:correlated_noise}(c), we plot the deformation estimation error at different resolutions. Comparing these plots with those for noiseless data in Fig.~\ref{fig:model_mismatch}, we see that the errors at the boundaries are higher for noisy data, which is most clearly observed at the coarse resolutions.

\paragraph{Deformation estimation is limited by the model basis} We performed experiments with realistic 3D simulated data where the ground truth deformation field along the $z$ direction contained cubic terms in $x$ and $y$ in addition to the quadratic terms in \eqref{eq:TEM_sim_gt_defo}. The ground truth deformation field used in these experiments was given by:
\begin{equation}
    D_z(x, y, z, t) = (P_0 + P_1 x^2 + P_2 y^2 + P_3 xy^2 + P_4 x^2y) t
\end{equation}
with $P_0 = 200\ \text{nm}$, $P_1 = P_2 = -50\ \text{nm}^{-1}$, $P_3 = P_4 = 25\ \text{nm}^{-2}$.
Although the ground truth contained cubic terms, we restricted the deformation terms used in our forward model to be quadratic in $x$ and $y$. We performed experiments for both noiseless data and data corrupted with correlated Poisson noise. For both noiseless and noisy data, our algorithm was able to identify the quadratic terms in the deformation field (Fig.~\ref{fig:mismatch_defo_field}(a-b)). As there were no cubic terms in the forward model, the reconstructed deformation fields did not contain any cubic components. The effect of this mismatch is greatest at the two corners of the FoV where the contribution of cubic terms was the highest. 

When we included cubic terms in the forward model, we found that both marker localization and deformation estimation improved as both quadratic and cubic terms were now estimated. The recovered deformation field in Fig.~\ref{fig:model_mismatch}(c) is much closer to the ground truth. These results indicate that the accuracy of SparseAlign is limited by the basis used for deformation modelling.

\begin{figure*}
    \centering
    \includegraphics[scale=0.5]{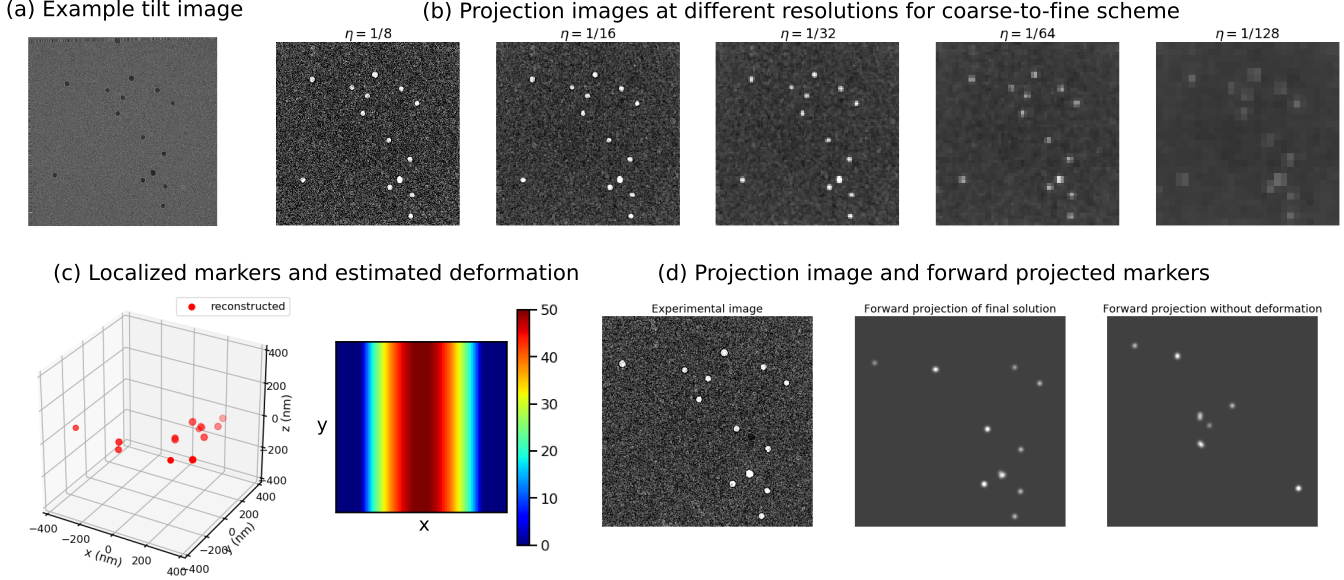}
    \caption{Results on experimental data. (a) A raw projection image in the acquired tilt series. (b) One image from pre-processed data used for deformation estimation and marker localization with downsampling factor $\eta = 1/8, 1/16, 1/32, 1/64, 1/128$. (c) Localized markers (left) and estimated deformation along $z$ (in nm). (d) One experimental projection image downsampled by $\eta=1/8$ (left), forward projection of localized markers with estimated deformation field (centre) and forward projection of markers with deformation field set to zero (right).}
    \label{fig:real_data}
\end{figure*}

\paragraph{SparseAlign locates markers reasonably in experimental data} We used an experimental dataset of gold beads embedded in ice to test the applicability of our method to experimental datasets. We used a coarse-to-fine scheme with downsampling factors $\eta = 1/128, 1/64, 1/32, 1/16, 1/8$ to localize gold bead markers and estimate the deformation field. We show an example tilt image in Fig.~\ref{fig:real_data}(a) and the same image at different downsampling factors in Fig.~\ref{fig:real_data}(b). 

Using a coarse-to-fine scheme we were able to localize several, but not all, markers. In Fig.~\ref{fig:real_data}(c), we show our marker localization results. We thresholded the localized markers according to their reconstructed weights. Here we show 15 markers with the highest weights. We estimated deformation along the $z$ direction using a quadratic model:
\begin{equation}
    D_{t,z}(r,P) = (P_{0} + P_{1}x + P_{2}y + P_{3}x^2 + P_{4}y^2 + P_{5}xy)t
\end{equation}
Additionally, we set the $x$ and $y$ components of the deformation field to zero. It is probable that our assumed deformation field was insufficient to model sample deformation in the experimental data. 

Our algorithm predicted a deformation field that is quadratic in $x$ but constant in $y$, a model that could not be experimentally validated. Plugging the estimated deformation field and marker locations into our forward model, we computed the forward projection shown in Fig.~\ref{fig:real_data}(d). Comparing this image to the data, we see that not all markers have been localized correctly, but at least one marker was localized in each of location with a cluster of markers. Markers throughout the FoV were localized; this suggests that the deformation estimation routine did not do worse for certain spatial regions. Moreover, mismatch in the shapes of actual markers and the Gaussian used in our forward model did not hinder the localization of most markers. Using localized marker locations and setting deformation to zero leads to projection images that are qualitatively different from the experimental data (Fig.~\ref{fig:real_data}(d)).

\section{Conclusion and Discussion}\label{conclusion}
Marker-based alignment is a important step for reconstruction improvement in cryoET. We have developed a mathematical approach called SparseAlign for compensating beam-induced local sample motion. In contrast to current approaches, our method does not need labelled marker locations, and directly uses projection data to localize markers and solve for the parameters of a polynomial deformation model. As a consequence, our method is more suited to data with low signal-to-noise ratios where markers cannot be reliably identified. 

Despite solving a more ill-posed problem for deformation estimation, SparseAlign localizes markers and estimates deformation parameters with an accuracy comparable to that of the doming model approach. Moreover, SparseAlign estimates deformation accurately even when the forward model for markers shows discrepancies with respect to marker projections in observed data.  

The image-based loss \eqref{eq:adcg_opt} in this paper can be improved upon by using a more canonical loss as the objective function for marker localization and deformation estimation. Unlike the $\ell^2$ loss used in this paper, the Wasserstein loss measures distances between distributions and has non-zero gradients even when the supports of the ground truth and current solution do not overlap \cite{kolouri2017optimal}.

The application of our approach to experimental data is limited by the deformation model used. One way to choose the most suitable, sparse basis for deformation modelling is to optimize over a library of basis functions using the data-driven approach in \cite{brunton2016discovering}.  

In this paper, we have ignored the image contrast of the biological sample while estimating deformation parameters. Ideally, our approach would be the first step in an iterative scheme where we alternate between sample reconstruction and tilt-series alignment, taking both sample and marker contributions into account during deformation estimation.

\section*{Acknowledgments}
This work was supported by the European Union's Marie Skłodowska-Curie Innovative Training Network MUMMERING (grant
agreement no. 765604) and the Netherlands Organisation for Scientific Research (NWO) (project number 613.009.106). E.F.~thanks Felix de Haas, Sebastian Unger, and Oliver Raschdorf for their help with sample preparation and experimental data collection. P.S.G.~thanks Roberto Bondesan for fruitful discussions.

\bibliographystyle{IEEEtran}
\bibliography{IEEEabrv,bibliography}

\end{document}


\maketitle

\section{Implementation Notes}\label{impl-notes}
We implemented SparseAlign in Python 3.6 and used several Python packages for each subroutine. Marker location refinement was performed using automatic differentiation routines in \texttt{Autograd} \cite{maclaurin2015autograd} and the L-BFGS-B method implemented in \texttt{SciPy} \cite{virtanen2020scipy}. The L-BFGS-B routine in \texttt{SciPy} was also  used for deformation estimation. 

We generated Gaussian noise and Poisson noise for 3D simulated data using \texttt{NumPy} \cite{harris2020array}. Segmentation using Otsu's method was performed for 3D simulated data using pre-defined functions in \texttt{scikit-image} \cite{van2014scikit}. 

All experiments were performed using JupyterLab notebooks \cite{kluyver2016jupyter} on an Intel(R) Core(TM) i7-8700K CPU with 12 cores. Example code for 2D alignment using SparseAlign is available in the GitHub repository \url{https://github.com/poulamisganguly/SparseAlign/}.

\begin{figure*}
    \centering
    \includegraphics[scale=0.45]{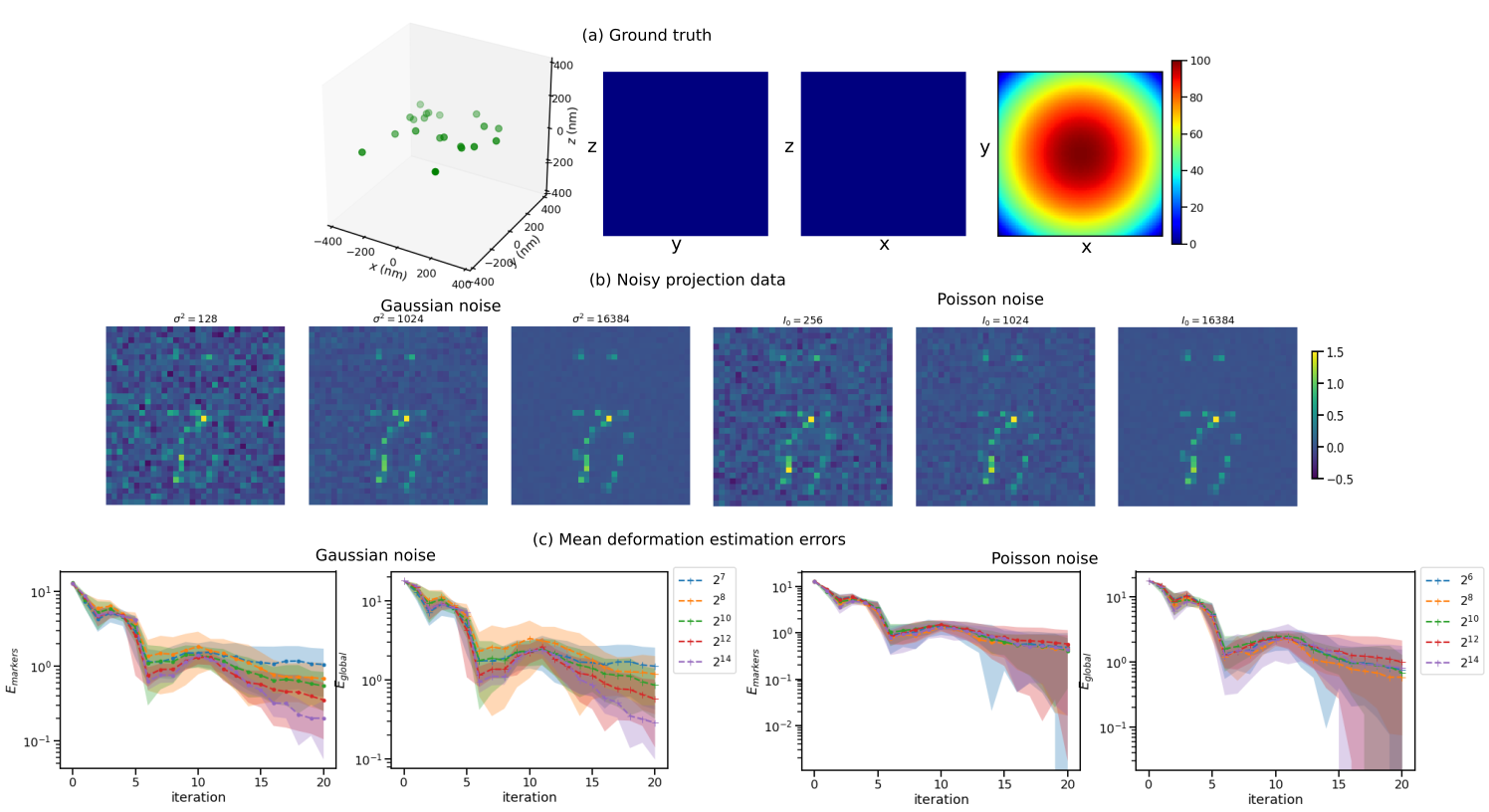}
    \caption{Deformation estimation in 3D with Gaussian and Poisson noise-corrupted data. (a) Ground truth configuration of markers (left) and ground truth deformation field in nm (right). (b) Projection image at $0^\circ$ with different Gaussian noise and Poisson noise settings. The variance of Gaussian noise ($\sigma^2$) and the photon flux ($I_0$) were chosen to simulate comparable Gaussian noise and Poisson noise realizations. (c) Deformation field estimation errors as a function of iteration at markers ($E_\text{markers}$) and in the entire field-of-view ($E_\text{global}$) for various Gaussian and Poisson noise settings.}
    \label{fig:effect_of_noise}
\end{figure*}

\begin{figure*}
    \centering
    \includegraphics[scale=0.45]{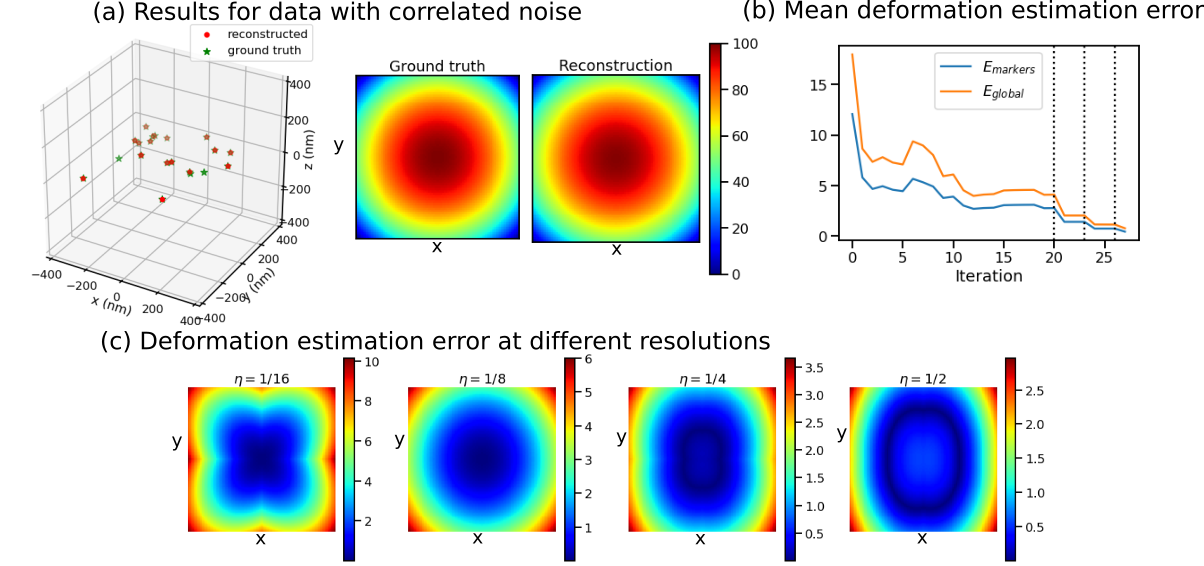}
    \caption{Results on realistic marker data with correlated noise. (a) Marker localization (left) and deformation estimation (right) results on data corrupted with realistic correlated noise. Deformation values shown in nm.
    (b) Mean deformation estimation error at ground truth marker locations and in the entire field-of-view for each iteration. Resolution changes in the coarse-to-fine scheme used to solve for marker locations and deformation parameters are indicated with black dotted lines. (c) Absolute deformation estimation error (along $z$) with respect to the ground truth at different values of the downsampling factor $\eta$.}
    \label{fig:correlated_noise}
\end{figure*}

\begin{figure*}
    \centering
    \includegraphics[scale=0.5]{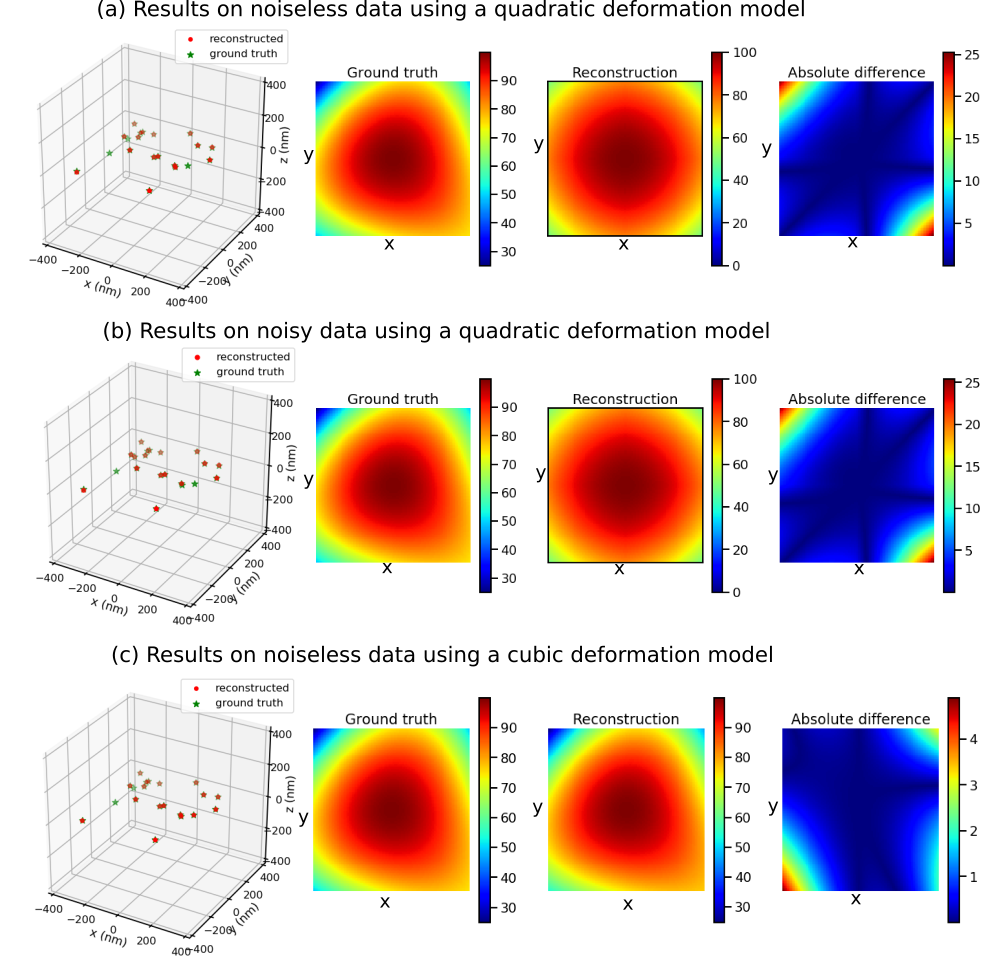}
    \caption{Alignment using a mismatched deformation model. Marker localization and deformation estimation for (a) noiseless and (b) noisy data using a quadratic deformation field model and a cubic deformation field as ground truth. (c) Marker localization and deformation field estimation using a cubic deformation field model for noiseless data. All deformation values shown in nm.}
    \label{fig:mismatch_defo_field}
\end{figure*}

\bibliographystyle{IEEEtran}
\bibliography{IEEEabrv,bibliography}